\documentclass[prb,twocolumn,showpacs,floatfix]{revtex4}
\usepackage{graphicx}
\usepackage{dcolumn}
\usepackage{bm}

\begin{document}
\title{Wave packet dynamics in hole Luttinger systems}

\author{V.~Ya.~Demikhovskii, G.~M.~Maksimova, and E.~V.~Frolova}
\email{demi@phys.unn.ru} \affiliation{Nizhny Novgorod State
University,\\ Gagarin Ave., 23, Nizhny Novgorod 603950, Russian
Federation}

\date{today}

\begin{abstract}
For hole systems with an effective spin $3/2$ we analyzed
analytically and numerically the evolution of  wave packets with
the different initial polarizations. The dynamics of such systems
is determined by the $4\times 4$ Luttinger Hamiltonian.  We work
in the space of arbitrary superposition of light- and heavy-hole
states of the "one-particle system".  For 2D packets we obtained
the analytical solution for the components of wave function and
analyzed the space-time dependence of probability densities as
well  as angular momentum densities. Depending on the value of the
parameter $a=k_0d$ ($k_0$ is the average momentum vector and $d$
is the packet width) two scenarios of evolution are realized. For
$a>>1$ the initial wave packet splits into two parts and the
coordinates of packet center experience the transient oscillations
or {\it Zitterbewegung} (ZB) as for other two-band systems. In the
case when $a<<1$ the distribution of probability density at $t>0$
remains almost cylindrically symmetric and the ripples arise at
the circumference of wave packet. The ZB in this case is absent.
We evaluated and visualized for different values of parameter $a$
the space-time dependence of angular momentum densities, which
have the multipole structure. It was shown that the average
momentum components can precess  in the absence of external or
effective magnetic fields due to the interference of the light-
and heavy hole states. For localized initial states this
precession has a transient character.
\end{abstract}


\maketitle

\section{Introduction}

In a crystalline solids described by the Hamiltonian
$\hat{H}=\hat{\vec{p}}^2/2m+V(\vec{r})$, where $\hat{\vec p}$ is
the momentum operator, $V(\vec{r})$ is a periodic potential, and
$m$ is the electron mass in vacuum, the electron velocity operator
$\hat{\vec{v}}=\hat{\vec{p}}/{m}$ does not commute with the
Hamiltonian $\hat H$ and so, the velocity is not a constant of
motion. As a result, various approximations including nearly-free,
tightly bound electron models or $\vec k \cdot \vec p$ methods
lead to unusual dynamics of charged particles, which is completely
different from their average behavior. For the motion of electrons
in solids, a highly oscillatory component appears in the time
evolution of physical observables such as position, velocity and
spin angular momentum. This phenomenon which is known as {\it
Zitterbewegung} (ZB) was described by Schr\"{o}dinger in 1930 for
a free relativistic electron in vacuum.\cite{SchBarut, Lok} It was
later understood that the ZB (a trembling motion) is due to the
interference of states with positive and negative electron
energies.

At first the ZB phenomena in crystalline solids was predicted with
the use of LCAO method in Refs.[3-5]. Later an oscillatory motion
of electron wave packets has been considered in wide class of 3D
solids and nanostructures, including narrow gap
semiconductors\cite{Zaw72}, carbon nanotubes\cite{Zaw74}, 2D
electron gas with Rashba spin-orbit coupling\cite{SchLosWest,DMF},
 single and bilayer graphene\cite{Katsn,MDF}, and also
superconductors\cite{Lur}. The ZB of photons near the Dirac point
in 2D photonic crystals was discussed in Ref[13]. Rusin and
Zavadski considered ZB in crystalline solids for nearly-free and
tightly bound electrons and concluded that ZB in solids is a rule
rather than an exception.\cite{RusZaw19} They determined the
parameters of trembling motion and concluded that, when the bands
are decoupled, electrons should be treated as particles of a
finite size. In another work by Rusin and Zavadski was performed
the calculations of ZB in the Kronig-Penney model and was
demonstrated that the two-band $(\vec{k}\cdot\vec{p})$ model is
adequate for this description.\cite{RusZaw}

The authors of Ref. [16] studied the semiclassical motion of holes
by numerical solution of equations for Heisenberg operators in the
Luttinger model at the presence of dc electric field. The
trajectories and angular momentum oscillations of heavy and light
holes reminiscent of the ZB were found and analyzed.

In Refs. [17,18] was presented theoretical analysis of ZB for
systems characterized by different Hamiltonians with gapped or
spin-orbit coupled spectrum, including Luttinger, Kane, Rashba and
Dresselhaus. It was demonstrated the analogy of ZB in all these
systems and presented a unified treatment of these phenomena.
Culcer {\it et al.} showed that in the Luttinger model the time
dependence of the angular momentum operator $\hat{\vec{S}}(t)$
corresponds to a spin precession in the absence of any external or
effective magnetic field.\cite{CulWin} In Ref. [20] the formalism
for treating the wave packet evolution in solids with two band
energy spectrum has been developed, where the nontrivial
non-Abelian terms in the equation of motion, arising from the
additional degree of freedom, were obtained. In particular, the
effect of spin separation in electric field was analyzed.

Mainly the ZB like effects were considered in the frame of
Heisenberg representation. In this representation the time
evolution of the system can be inferred directly from the
differential equations that governed the behavior of operators
$\hat{\vec{r}}(t)$, $\hat{\vec{v}}(t)$ and $\hat{\vec{S}}(t)$. As
usual these studies have dealt with the behavior of expectation
values of these operators without going into the details of the
space-time evolution.

In this work we study the space-time evolution of the wave packets
in Schr\"{o}dinger representation for the Luttinger model which
describes the states of heavy and light holes in vicinity of the
top of energy band in a wide class of $p$ - doped III-V
semiconductors. We obtain not only the average values such as
average velocities or average angular momentum but more
informative characteristics of hole wave packets: the probability
density and effective spin densities.We demonstrate that
space-time dynamics of wave packets has, as a rule, a complex
character and it helps us to explain many interesting patterns of
evolution for different initial parameters and polarizations of
wave packets.

The paper is organized as follows. In section II the analytical
expressions for the $4$-component wave function are obtained for
different initial angular momentum polarizations. Two different
scenarios of the space-time evolution of probability densities for
different values of the parameter $a=k_0d$, where $k_0$ is average
momentum and $d$ is the initial packet width  are discussed. The
transient oscillations of the position operator are analyzed. The
effective spin dynamics is considered in Sec.III. We evaluate and
visualized the effective spin densities and average $\bar{\vec
S}(t)$. It was shown that the average angular momentum experiences
transient precession about the average packet momentum $\vec k_0$
when $\bar{\vec S}(0)$ is not perpendicular or parallel to $\vec
k_0$. We conclude with some remarks in Sec. IV.

\section{Space - time evolution and {\it Zitterbewegung}}

Holes near the top of valence band of common semiconductors such
as $Ge$ and $GaAs$ have an effective spin $j=3/2$ and are
described by the $4\times 4$ Luttinger Hamiltonian\cite{Lutt},
which in the isotropic approximation has the form\cite{Lipari}

$$\hat{H}=\frac{\hbar^2
}{2m}\Bigg[(-\gamma_{1}+\frac{5}{2}\gamma_{2})k^{2}-2\gamma_{2}(\hat{\vec{S}}\vec{k})^{2}\Bigg]
.\eqno(1)$$ Here $\gamma_{1}$ and $\gamma_{2}$ are Luttenger
parameters, $m$ is the electron mass in vacuum, and $\vec{S}$ is
the vector of $4\times4$ spin matrices correspondent to spin
$S=\frac{3}{2}$. The matrices $S_x$, $S_y$, $S_z$ in Eq.(1) are
given by

$$S_x=i\pmatrix{0 & \frac{\sqrt{3}}{2} & 0 & 0\cr
-\frac{\sqrt{3}}{2} & 0 & 1 & 0\cr 0 & -1 & 0 &
\frac{\sqrt{3}}{2}\cr 0 & 0 & -\frac{\sqrt{3}}{2} & 0 },$$

$$S_y=\pmatrix{0 & \frac{\sqrt{3}}{2} & 0 & 0\cr
\frac{\sqrt{3}}{2} & 0 & 1 & 0\cr 0 & 1 & 0 &
\frac{\sqrt{3}}{2}\cr 0 & 0 & \frac{\sqrt{3}}{2} & 0},$$

$$S_z=\pmatrix{\frac{3}{2} & 0 & 0 & 0\cr 0 & \frac{1}{2} & 0 &
0\cr 0 & 0 & -\frac{1}{2} & 0\cr 0 & 0 & 0 & -\frac{3}{2}}.$$

The two-fold degenerate energy eigenvalues of $\hat{H}$ for light
holes ($L$) and heavy holes ($H$) are

$$\displaylines{E_{L}(k)=\frac{\hbar^2
}{2m}(-\gamma_{1}+2\gamma_{2})k^{2},\cr\hfill
~~E_{H}(k)=-\frac{\hbar^2
}{2m}(\gamma_{1}+2\gamma_{2})k^{2}.\hfill\llap(2)\cr} $$

The helicity operator defined by
$\Lambda=\frac{\vec{k}\vec{S}}{k}$ commutes with Hamiltonian (1)
so that a helicity
$\lambda~(\lambda=\pm\frac{1}{2},\pm\frac{3}{2})$ is a good
quantum number: $\lambda=\pm\frac{1}{2}$ corresponds to the light
holes and $\lambda=\pm\frac{3}{2}$ to the heavy holes.

In this work we consider $2D$ model with $\vec{k}=(k_{x},k_{y})$
that allows us to obtain the simple analytical results concerning
the time-evolution of the initial wave packet $\Psi(x,y)$. In this
case the eigenstates of the helicity operator $\Psi_{\lambda}$ are

$$\displaylines{\Psi_{\lambda=\pm\frac{1}{2}}=\frac{1}{2\sqrt{2}}\pmatrix
{\pm i\sqrt{3}{\rm e}^{-i\varphi}\cr 1\cr \pm i{\rm
e}^{i\varphi}\cr \sqrt{3}{\rm
e}^{i2\varphi}},\cr\hfill\Psi_{\lambda=\pm\frac{3}{2}}=\frac{1}{2\sqrt{6}}\pmatrix
{\mp i\sqrt{3}{\rm e}^{-i\varphi}\cr -3\cr \pm i3{\rm
e}^{i\varphi}\cr \sqrt{3}{\rm e}^{i2\varphi}}.\hfill\cr}$$

Here we will use more simple eigenfunctions of the Hamiltonian (1)
which are the linear combinations of $\Psi_{\lambda}$

$$\Psi_{L(H),i}(\vec{r})=\varphi_{\vec{k}}(\vec{r})U_{L(H),i},~~~i=1,2,\eqno(3)$$
where $\varphi_{\vec{k}}(\vec{r})=\rm e^{i\vec{k}\vec{r}}/2\pi$
and spinors $U_{L(H),i}$ are given by

$$\displaylines{U_{L1}=\frac{1}{2}\pmatrix {0 \cr 1 \cr 0 \cr
\sqrt{3}{\rm e }^{i2\varphi}},~~~~U_{L2}=\frac{1}{2}\pmatrix
{\sqrt{3}{\rm e }^{-i2\varphi} \cr 0 \cr 1 \cr 0}\cr\hfill
U_{H1}=\frac{1}{2\sqrt{3}}\pmatrix {0 \cr -3 \cr 0 \cr
\sqrt{3}{\rm e }^{i2\varphi}},~U_{H2}=\frac{1}{2\sqrt{3}}\pmatrix
{\sqrt{3}{\rm e }^{-i2\varphi} \cr 0 \cr -3 \cr 0}
\hfill\llap{(4)}\cr}$$ with $k_{x}=k\cos\varphi$,
$k_{y}=k\sin\varphi$.

Now the most general wave function can be written as

$$\displaylines{\Psi(\vec{r},t)=\sum\limits_{i=1,2}\int
d\vec{k}\varphi_{\vec{k}}(\vec{r})\Bigg(C_{L,i}(\vec{k})U_{L,i}{\rm
e }^{-iE_{L}t/\hbar}+\cr\hfill +C_{H,i}(\vec{k})U_{H,i}{\rm e
}^{-iE_{H}t/\hbar}\Bigg),\hfill\llap(5)\cr}$$ where
$C_{L(H),i}(\vec{k})$ are to be determined from the Fourier
expansion of $\Psi(\vec{r})$ at $t=0$.

i) As a first example we compute the time evolution of the
Gaussian spinor

$$\Psi(\vec{r},0)=\frac{1}{d\sqrt{\pi}}\exp(-\frac{r^2}{2d^2}+ik_{0}x)\pmatrix{1\cr
0\cr 0\cr 0}.\eqno(6)$$ It realizes a wave packet with positive
average momentum and the effective spin oriented initially along
$z$ axis, $S_z=\frac{3}{2}$. The appropriate coefficients can be
readily found from Eq.(5)

$$C_{L1}=C_{H1}=0,~C_{L2}=\frac{\sqrt{3}}{2}f_{\vec{k}}{\rm
e}^{i2\varphi},~C_{H2}=\frac{1}{2}f_{\vec{k}}{\rm
e}^{i2\varphi},\eqno(7)$$ where $f_{\vec{k}}$ is the Fourier
transform of wave packet (6)

$$f_{\vec{k}}=\frac{d}{\sqrt{\pi}}\exp\Bigg(-\frac{(k_{x}-k_{0})^2d^2}{2}-\frac{k_{y}^2d^2}{2}\Bigg).\eqno(8)$$
As it follows from Eq.(7), the initial wave packet consists of the
light-hole and heavy- hole states, and the weight of the light
hole states is three time greater than that of heavy holes.

Substituting Eqs.(7), (8) into Eq.(5) and performing integration
we finally have the components of wave function
$\Psi(\vec{r},t)=(\Psi_{1}(\vec{r},t),\Psi_{2}(\vec{r},t),
\Psi_{3}(\vec{r},t),\Psi_{4}(\vec{r},t))^T$:

$$\Psi_{2}(\vec{r},t)=\Psi_{4}(\vec{r},t)=0,\eqno(9)$$

$$\Psi_{1}(\vec{r},t)=\frac{d{\rm
e}^{-a^2/2}}{4\sqrt{\pi}}\Bigg[\frac{3{\rm
e}^{\beta/2\delta_{L}}}{\delta_{L}}+ \frac{{\rm
e}^{\beta/2\delta_{H}}}{\delta_{H}}\Bigg],\eqno(10)$$

$$\displaylines{\Psi_{3}(\vec{r},t)=\frac{\sqrt{3}d{\rm
e}^{-a^2/2}}{4\sqrt{\pi}}\frac{a^2d^2+(x+iy)^2}{(ad+y)^2+x^2}\times\cr\hfill\times
\Bigg[\frac{{\rm
e}^{\beta/2\delta_{L}}}{\delta_{L}}(1-\frac{2\delta_{L}}{\beta})-
\frac{{\rm
e}^{\beta/2\delta_{H}}}{\delta_{H}}(1-\frac{2\delta_{H}}{\beta})\Bigg],\hfill\llap{(11)}\cr}
$$ where

$$\displaylines{\delta_{L}=d^2+i\hbar(-\gamma_{1}+2\gamma_{2})t/m,\cr\hfill
\delta_{H}=d^2-i\hbar(\gamma_{1}+2\gamma_{2})t/m,\hfill \cr\hfill
\beta=(ad+ix)^2-y^2.\hfill\llap(12)\cr}$$

Two terms in Eq.(10), (11) labelled by $L$ and $H$ indices show
the contribution of light and heavy holes correspondingly. In
particular the item $1/\delta_L\exp(\beta/2\delta_L-a^2/2)$ in
Eq.(10), which can be written as

$$\displaylines{\frac{\exp(\beta/2\delta_L-a^2/2)}{\delta_L}=
\cr\hfill \exp\Bigg(\frac{-(x^2+y^2)+2iad^3(x+\frac{\hbar k_0}
{2m}(-\gamma_1+2\gamma_2))}{2(d^2+i\hbar(-\gamma_1+2\gamma_2)t/m)}\Bigg)\times\hfill\cr\hfill
\times (d^2+i\hbar(-\gamma_1+2\gamma_2)t/m)^{-1},\hfill\cr}$$
reproduces the time evolution of Gaussian wave packet for spinless
particle with energy $E_L=\frac{\hbar^2
}{2m}(-\gamma_1+2\gamma_2)k^2$.

\begin{figure}
  \centering
  \includegraphics[width=70mm]{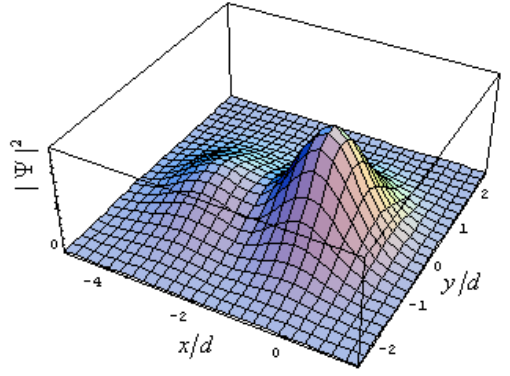}
\caption{(Color online) The full hole density for initial wave
packet, Eq.(6) at the moment $t=0.2$ (in the units of $t_0=\frac{m
d^2}{2\gamma_2\hbar}$) for $ \gamma_1=7.65$ and $\gamma_2=2.41$,
for width $d=3\cdot10^{-6} cm$ and $k_0=2\cdot10^6
cm^{-1}$.}\label{Fig1}
\end{figure}

\begin{figure}
  \centering
  \includegraphics[width=70mm]{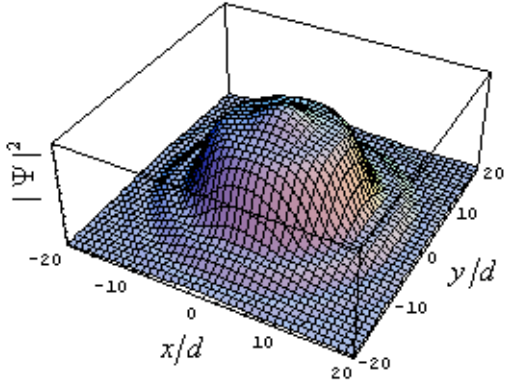}
\caption{(Color online) The full hole density at the moment $t=4$
(in the units of $t_0$) for $\gamma_1=7.65$ and $\gamma_2=2.41$
for initial wave packet, Eq.(6) with width $d=3\cdot 10^{-6} cm$
and $k_0=0$.}\label{Fig2}
\end{figure}

In Fig.1 and Fig.2 we represent the hole density for initial wave
packet, Eq.(6) for $\gamma_1=7.65$ and $\gamma_2=2.41$, which
correspond to $GaAs$, with width $d=3\cdot 10^{-6}cm$ and
$k_0=2\cdot 10^{6}cm^{-1}$ at the moment $t=0.2$ (in units of
$t_0=m d^2/2\gamma_2\hbar$) (Fig.1); with width $d=3\cdot
10^{-6}cm$ and $k_0=0$  at the moment $t=4$ (in units of $t_0$)
(Fig.2).

It is not difficult to see in Fig.1 and Fig.2 that the parameter
$a=k_0d$ regulates the character of wave packet evolution. For the
case $a>>1$ (see Fig.1) the initial wave packets split into two
parts and their evolution is accompanied by the {\it
Zitterbewegung} phenomenon. This dynamics is a result of
interference of the hole states laying near the point $\vec k_0$
in momentum space. The splitting of the wave packets into two
parts which propagate with unequal group velocity and have
different angular momentum polarizations, appears due to the
presence of the light and heavy holes states in the expansion the
initial wave packet. As usual, the splitting is accompanied by the
packet broadening which appears due to the effect of dispersion.

The ZB is a result of interference and a complex space-time
evolution of two parts of wave packet. In another extreme case,
when the inequality $a<<1$ takes place, the oscillations of the
probability density at the circumference of the wave packet are
closely related to the interference of heavy- and light-hole
states located in vicinity of the point $\vec k=0$ in the momentum
space (see Fig.2). This interference gives rise to the formation
of the circular ripples around the packet center. The local period
of these oscillations of probability density is not constant, it
decreases with increasing radius of the ripple and time. This
conclusion can be illustrated by the analytical expression for the
density probabilities for $a=0$. For example, the analytical
expression for the first component of wave function is

$$\displaylines{|\Psi_1(\vec r,t)|^2=\frac{1}{16\pi
d^2}\Bigg\{\frac{9\exp(-\frac{r^2}{d^2(1+\omega_L^2t^2)})}{1+
\omega_L^2t^2}+\frac{\exp(-\frac{r^2}{d^2(1+\omega_H^2t^2)})}{1+\omega_H^2t^2}\cr\hfill
+6\frac{\exp(-\frac{r^2}{2d^2(1+\omega_L^2t^2)}
-\frac{r^2}{2d^2(1+\omega_H^2t^2)})}{1+\omega_L^2\omega_H^2t^4+\omega_L^2t^2+\omega_H^2t^2}\Bigg[(1+\omega_L\omega_Ht^2)\times
\hfill\cr\hfill \times
\cos(\xi(t,r))+(\omega_L-\omega_H)t\sin(\xi(t,r))\Bigg]\Bigg\},\hfill\cr}$$
where $\omega_L=\hbar(-\gamma_1+2\gamma_2)/d^2m$,
$\omega_H=-\hbar(\gamma_1+2\gamma_2)/d^2m$,
$\Omega=\frac{\omega_L}{1+\omega_L^2t^2}+\frac{\omega_H}{1+\omega_H^2t^2}$,
and $\xi(t,r)=\frac{r^2t}{2d^2}\Omega$. This expression contains
oscillating terms which are proportional to functions
$\cos(\xi(t,r))$ and $\sin(\xi(t,r))$, describing the ripples
structure at Fig.2. It should be noted, however, that these
oscillations do not provoke the ZB of average coordinate of packet
center.

\begin{figure}
  \centering
  \includegraphics[width=60mm]{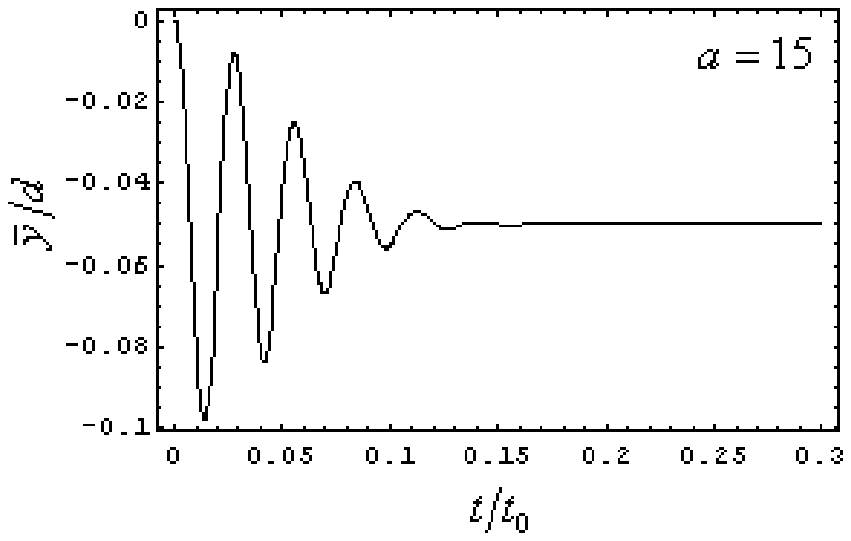}
  \includegraphics[width=60mm]{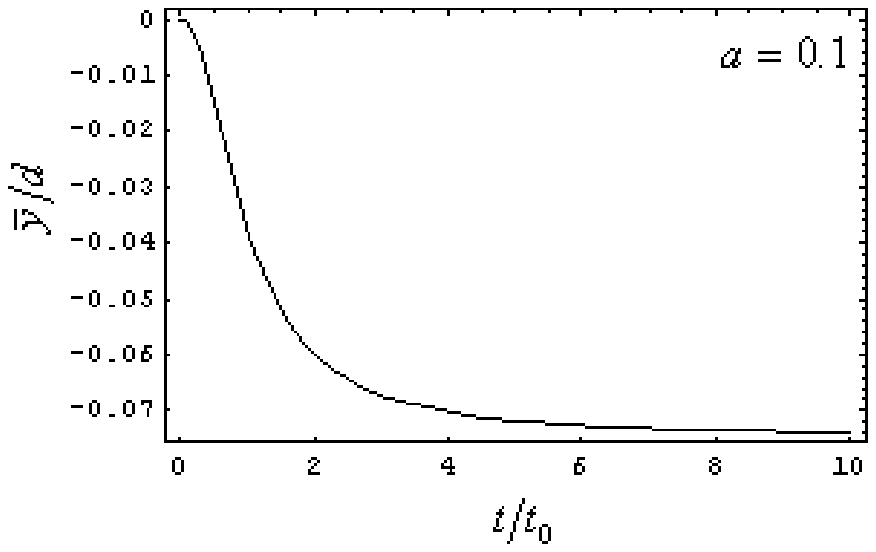}
\caption{(Color online) The average coordinate $\bar y(t)$ of the
packet center for different cases of value $a=k_0d$.}\label{Fig3}
\end{figure}

To analyze the motion of the packet center we have to find the
average value of the position operator. To do it, let us use the
momentum representation. The components $C_{\alpha}(\vec{k},t)$
($\alpha=1,...4$) of wave function (5) in this representation can
be obtained from Eqs.(4), (5), (6). After that the usual
definition

$$\bar{x}_i(t)=\sum\limits_{\alpha=1}^{4}\int
d\vec{k}C_{\alpha}^\ast(\vec{k},t)i\frac{d}{dk_i}C_{\alpha}(\vec{k},t),~~i=1,2\eqno(13)$$
leads to

$$\bar{x}(t)=\frac{\hbar
k_0}{m}(\gamma_{2}-\gamma_{1})t,\eqno(14)$$

$$\bar{y}(t)=\frac{3}{4k_0}\Bigg[\exp(-\frac{(at/t_0)^2}{1+(t/t_0)^2})\cos(\frac{a^2t/t_0}{1+(t/t_0)^2})-1\Bigg].\eqno(15)$$

We see that the center of wave packet moves along $x$ direction
with constant velocity $\bar{V}_{0x}=\frac{\hbar k_0}{m}
(\gamma_{2}-\gamma_{1})$. This expression also can be found from
the equation

$$\bar{V}_{0x}=\frac{3}{4}V_{L}+\frac{1}{4}V_H,\eqno(16)$$ where
$V_{L(H)}=\frac{1}{\hbar}\frac{\partial E_{L(H)}}{\partial
k_x}|_{k_x=k_0}$ are the velocities of light (heavy) holes along
$x$ axis at $k_x=k_0$ and $E_{L(H)}$ are determined by Eq.(2). The
coefficients $3/4$ and $1/4$ correspond to relative parts of $L-$
and $H-$ holes. The motion of the wave packet center along $x$ is
accompanied by the oscillations of the packet center at $a>1$ in a
perpendicular direction or {\it Zitterbewegung} (Eq.(15)). It is
easy to see that such trembling motion has a transient character
as for other systems with two-band structures (Fig.3 for $a=15$).
As it follows from Eqs.(14), (15) for a given initial polarization
of the wave packet the ZB occurs in the direction perpendicular to
the initial momentum $p_{0x}=\hbar k_0$. At large enough time
$t>>t_0$ the packet center shifts in $y$ direction at the values
of
$$y_0=\frac{3}{4k_0}({\rm e}^{-a^2}-1)$$ in accordance with
Eq.(15). The Fig.3 for $a=0.1$ obviously demonstrates the
monotonic behavior of $\bar y(t)$.

ii) As a second example we compute the free space-time evolution
of the Gaussian packet

$$\Psi(\vec{r},0)=\frac{\exp
(-\frac{r^2}{2d^2}+ik_0x)}{d\sqrt{2\pi}}\pmatrix{1 \cr 0 \cr 1 \cr
0}.\eqno(17)$$

This expression is obtained as a superposition of two components,
corresponding to the light and heavy holes. It should be mentioned
that the relative weight of the light and heavy holes is not
constant in momentum space in contrast to the previous case,
Eq.(6). Really, as it follows from Eqs.(5), (17) the amplitudes
$C_{L(H),i}$ are

$$\displaylines{C_{L1}=C_{H1}=0,
~~~C_{L2}=\frac{\sqrt{2}}{4}f_{\vec{k}}(\sqrt{3}{\rm
e}^{i2\varphi}+1),\cr\hfill
C_{H2}=\frac{\sqrt{2}}{4}f_{\vec{k}}({\rm
e}^{i2\varphi}-\sqrt{3}).\hfill\llap(18)\cr}$$

Substituting Eq.(18) into Eq.(5) we obtain the components of wave
function

$$\Psi_2(\vec{r},t)=\Psi_4(\vec{r},t)=0,$$

$$\displaylines{\Psi_1(\vec{r},t)=\frac{\sqrt{2}}{8}\frac{d}{\sqrt{\pi}}{\rm
e}^{-a^2/2}\Bigg(\frac{3{\rm
e}^{\beta/2\delta_L}}{\delta_L}+\frac{{\rm
e}^{\beta/2\delta_H}}{\delta_H}\Bigg)+ \hfill\cr +
\frac{\sqrt{3}d{\rm e}^{-a^2/2}}{4 \sqrt{2
\pi}}\frac{a^2d^2+(x-iy)^2}{(ad-y)^2+x^2}\Bigg(\frac{{\rm
e}^{\beta/2\delta_{L}}}{\delta_{L}}(1-~~~~\cr\hfill-\frac{2\delta_{L}}{\beta})-
\frac{{\rm
e}^{\beta/2\delta_{H}}}{\delta_{H}}(1-\frac{2\delta_{H}}{\beta})\Bigg),\hfill\llap(19)\cr}$$

$$\displaylines{\Psi_3(\vec{r},t)=\frac{\sqrt{2}}{8}\frac{d}{\sqrt{\pi}}{\rm
e}^{-a^2/2}\Bigg(\frac{{\rm
e}^{\beta/2\delta_L}}{\delta_L}+\frac{{3\rm
e}^{\beta/2\delta_H}}{\delta_H}\Bigg)+ \hfill\cr +
\frac{\sqrt{3}d{\rm
e}^{-a^2/2}}{4\sqrt{2\pi}}\frac{a^2d^2+(x+iy)^2}{(ad+y)^2+x^2}\Bigg(\frac{{\rm
e}^{\beta/2\delta_{L}}}{\delta_{L}}(1-~~~~\cr\hfill-\frac{2\delta_{L}}{\beta})-
\frac{{\rm
e}^{\beta/2\delta_{H}}}{\delta_{H}}(1-\frac{2\delta_{H}}{\beta})\Bigg),\hfill\llap(20)\cr}$$
where $\delta_L$, $\delta_H$ and $\beta$ are determined by
Eq.(12). The results of the calculation of average values of $x$
and $y$ for this polarization are

$$\bar x(t)=V_{0x}t+x_{ZB}(t),\eqno(21)$$ where the constant
velocity is defined by the expression

$$V_{0x}=\frac{\hbar
k_0}{m}\Bigg[-\gamma_1+\gamma_2\sqrt{3}\Bigg(1-\frac{1}{a^2}+\frac{1}{a^4}-\frac{{\rm
e}^{-a^2}}{a^4}\Bigg)\Bigg],\eqno(22)$$ and $x_{ZB}(t)$ describes
the oscillatory motion ({\it Zittebewegung})

$$\displaylines{\bar
x_{ZB}(t)=\frac{\sqrt{3}}{2k_0a^2}\exp(-\frac{a^2t^2/t_0^2}{1+t^2/t_0^2})\Bigg(\sin(\frac{a^2t/t_0}{1+t^2/t_0^2})-
\frac{t}{t_0}\times~~~~\cr\hfill \times
\cos(\frac{a^2t/t_0}{1+t^2/t_0^2})\Bigg)+\frac{\sqrt{3}}{2k_0a^2}{\rm
e }^{-a^2}\frac{t}{t_0},\hfill\llap(23)\cr}$$

$$\displaylines{\bar{y}(t)=\frac{\sqrt{3}}{4k_0}\Bigg[1-\frac{1}{a^2}-(1-\frac{1}{a^2})\exp(-\frac{a^2t^2/t_0^2}{1+t^2/t_0^2})\cos(\frac{a^2t/t_0}{1+t^2/t_0^2})+\cr\hfill
+
\frac{t}{a^2t_0}\exp(-\frac{a^2t^2/t_0^2}{1+t^2/t_0^2})\sin(\frac{a^2t/t_0}{1+t^2/t_0^2})\Bigg].\hfill\llap(24)\cr}$$

The space-time diagram of this solution, shown in Fig.4 again
shows the {\it Zitterbewegung} of the position's mean values. It
is interesting to stress that for the packet polarization
determined by Eq.(17) the oscillatory behavior occurs in both $x$
and $y$ directions. The amplitude of oscillations in $x$ direction
is much less than that in $y$ direction.

\begin{figure}
  \centering
  \includegraphics[width=60mm]{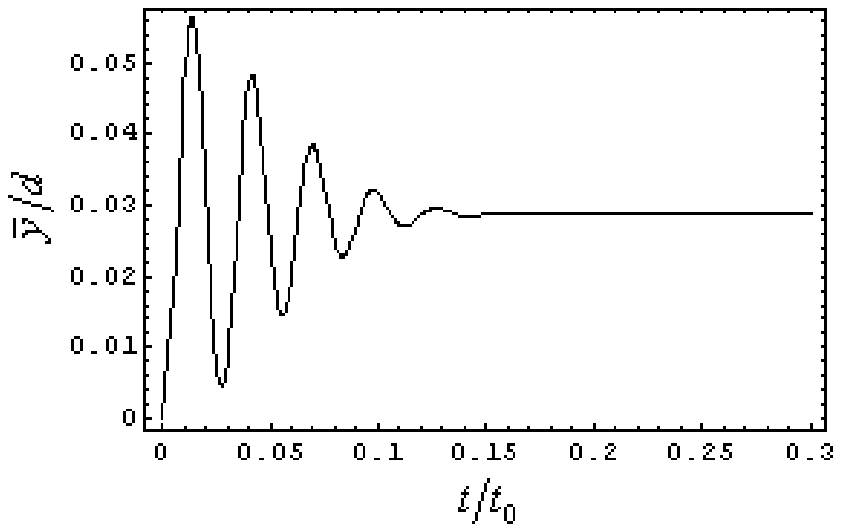}
  \includegraphics[width=60mm]{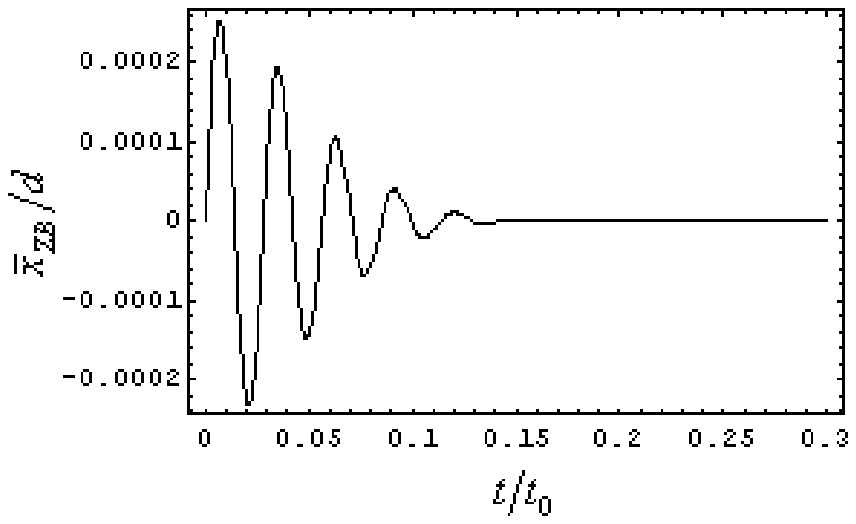}
\caption{(Color online) The space-time diagram of the packet
center for the initial wave packet, Eq. (17) with
$a=15$.}\label{Fig4}
\end{figure}

As it follows from Eqs.(15), (23), (24) the character of trembling
motion strongly depends on the parameter $a=k_0d$ as for the other
system.\cite{DMF,MDF} Really, it is easy to see that oscillations
of ${\bar {\vec r}}(t)$ (which are connected with the terms
$\cos(\frac{a^2t/t_0}{1+t^2/t_0^2})$ or
$\sin(\frac{a^2t/t_0}{1+t^2/t_0^2})$) occur if $a^2 \geq \pi$.
However the value $a$ should not be too large since the amplitude
of oscillations ($\sim \exp(-\frac{a^2t^2/t_0^2}{1+t^2/t_0^2})$)
decreases rapidly when $a$ increase.

Notice also that if we consider the plane wave that corresponds to
$d\rightarrow\infty$ (i.e. $a\rightarrow\infty$ and
$t_0\rightarrow\infty$) then we obtain from Eqs.(14), (15)

$$\displaylines{\bar x(t)=\frac{\hbar
k_0}{m}(\gamma_2-\gamma_1)t,\cr \hfill \bar
y(t)=\frac{3}{4k_0}\Bigg(\cos(\frac{2\gamma_2\hbar
k_0^2t}{m})-1\Bigg),\hfill \llap(25)\cr}$$ and correspondingly for
the second polarization (Eq.(17))

$$\displaylines{\bar{x}(t)=\frac{\hbar
k_0}{m}(\sqrt{3}\gamma_2-\gamma_1)t,\cr \hfill \bar y(t)=
\frac{\sqrt{3}}{4k_0}\Bigg(1-\cos(\frac{2\gamma_2\hbar
k_0^2t}{m})\Bigg).\hfill \llap(26)\cr}$$

iii) Let us consider now the third example when the initial wave
packet has the following form:

$$\Psi(\vec{r},0)=\frac{\exp
(-\frac{r^2}{2d^2}+ik_0x)}{d\sqrt{2\pi}}\pmatrix{1 \cr i \cr 0 \cr
0}.\eqno(27)$$

Using expression (5) and specifying the coefficients
$C_{L(H),i}(\vec{k})$, we find the components of the wave function
at arbitrary time:

$$\Psi_{1}(\vec{r},t)=\frac{d{\rm
e}^{-a^2/2}}{4\sqrt{2\pi}}\Bigg[\frac{3{\rm
e}^{\beta/2\delta_{L}}}{\delta_{L}}+ \frac{{\rm
e}^{\beta/2\delta_{H}}}{\delta_{H}}\Bigg],\eqno(28)$$

$$\Psi_{2}(\vec{r},t)=i\frac{d{\rm
e}^{-a^2/2}}{4\sqrt{2\pi}}\Bigg[\frac{{\rm
e}^{\beta/2\delta_{L}}}{\delta_{L}}+ \frac{3{\rm
e}^{\beta/2\delta_{H}}}{\delta_{H}}\Bigg],\eqno(29)$$

$$\displaylines{\Psi_{3}(\vec{r},t)=\frac{\sqrt{3}d}{4\sqrt{2\pi}}{\rm
e}^{-a^2/2}\frac{a^2d^2+(x+iy)^2}{(a^2d+y)^2+x^2}\times\cr\hfill\times
\Bigg[\frac{{\rm
e}^{\beta/2\delta_{L}}}{\delta_{L}}(1-\frac{2\delta_{L}}{\beta})-
\frac{{\rm
e}^{\beta/2\delta_{H}}}{\delta_{H}}(1-\frac{2\delta_{H}}{\beta})\Bigg],\hfill\llap{(30)}\cr}
$$

$$\Psi_{4}(\vec{r},t)=i\Psi_{3}(\vec{r},t)\eqno(31)$$

It is easy to see that for this polarization the time dependence
of the full probability density is similar to the evolution in the
case i) in sprite of that here all four components of wave
function are not equal to zero. Here we note only that the average
coordinate $\bar x(t)=-\frac{\hbar k_0\gamma_1}{m}t$, and $\bar
y(t)$ has the same form as in the case i), Eq.(15). But as we will
see below the spin density and average spin evolution in this case
differ qualitatively from those in cases i) and ii).

\section{Alternating spin - 3/2 polarization}

As was shown firstly in Ref.[19], the spin dynamics of spin - 3/2
hole is qualitatively different from those of spin - 1/2 electron
systems. In contrast to spin - 1/2 electron systems, neither the
magnitude nor the orientation of the angular momentum $\bar {\vec
S}$ are conserved. In spite of the fact that the coupling of spin
and orbital degrees of freedom cannot be written in terms of
external or effective magnetic field, the hole spin polarization,
as well as the higher-order multipoles demonstrate nontrivial
dynamics in time and can precess due to the interference of heavy-
and light- hole states\cite {CulWin}. In the Ref. [19] the main
results concerning the alternating spin polarization in spin - 3/2
hole systems were obtained only for the initial states represented
by the plane waves. Below we describe and visualize the spin
dynamics of hole wave packets with different initial
polarizations. For this purpose we consider the time evolution of
components of the effective spin density $S_i(\vec r,t)$ and their
average values $\bar{S_i}(t)$

$$S_i(\vec r,t)=\Psi^+(\vec r,t)\hat S_i \Psi(\vec r,t),
\eqno(32)$$

$$\bar S_i(\vec r,t)=\int\Psi^+(\vec r,t)\hat S_i \Psi(\vec
r,t)d\vec r, \eqno(33)$$

where $\Psi(\vec r,t)$ is a four-component wave function and
$\Psi^+(\vec r,t)$ denotes the Hermitian conjugated wave function.

Firstly, we consider the wave packet, corresponding to the hole
spin oriented initially along $z$ direction: $S_z=\frac{3}{2}$,
Eq.(6). In this case it is easy to see that spin densities are
equal to $S_x(\vec r,t)=S_y(\vec r,t)=0$, $S_z(\vec
r,t)=\frac{3}{2}|\Psi_1(\vec r,t)|^2-\frac{1}{2}|\Psi_3(\vec
r,t)|^2$, where the components of wave function $\Psi_1(\vec r,t)$
and $\Psi_3(\vec r,t)$ are determined by Eqs.(9)-(11).

To illustrate the evolution of the hole spin density $S_z(\vec
r,t)$ we plot this function in Fig.5 for the moment of the time
$t=15$ (in units of $t_0$) and for wave packet parameters
$d=10^{-6}$ $cm$ and $k_0=10^5$ $cm^{-1}$. Just as for hole
probability density the space-time evolution of the spin density
strongly depends on the parameter $a$. In particular, for small
values of $a$ as was demonstrated in Sec.II the splitting of wave
packet is absent and oscillations of the probability density
(ripples) arises on its periphery. Similar behavior of spin
density $S_z(\vec r,t)$ for $a=0.1$ one can observe in Fig.5.

Using Eq.(33) it is not difficult to find the average value $\bar
S_z(t)$ for given initial polarization

$$\displaylines{\bar{S_z}(t)=\frac{3}{4}\Bigg[1+\frac{\exp(-\frac{a^2(t/t_0)^2}{1+(t/t_0)^2})}{1+(t/t_0)^2}\times\cr\hfill
\times\Bigg(\cos(\frac{a^2t/t_0}{1+(t/t_0)^2})-
\frac{t}{t_0}\sin(\frac{a^2t/t_0}{1+(t/t_0)^2})\Bigg)\Bigg].\hfill\llap{(34)}\cr}$$

This expression obviously demonstrates that $\bar S_z(t)$
experiences the transient oscillations for the parameter $a>>1$
(see Fig.6). For $t/t_0>>1$ $\bar S_z(t)$ streams to the constant
values 3/4. Note that the length of spin vector in this case is
not constant. As it follows from Eq.(34) in the limiting case
$d\rightarrow \infty$, i.e. for the plane wave, $\bar S_z$ is

$$\bar S_z(t)=\frac{3}{2}\cos^2\frac{2\gamma_2\hbar
k_0^2t}{m}.\eqno(35)$$ This result coincides with that in
Ref.[19].

\begin{figure}
  \centering
  \includegraphics[width=70mm]{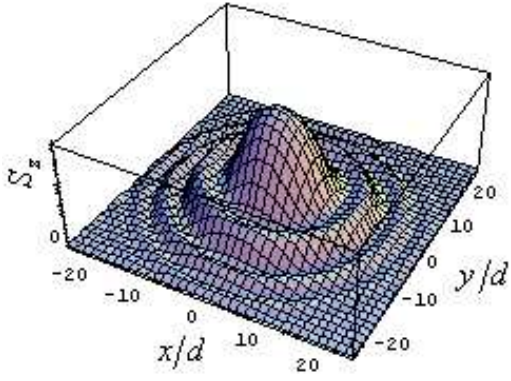}
\caption{(Color online) The angular momentum density $S_z(\vec
r,t)$ for initial wave packet, Eq.(6) with $k_0= 10^{5}~cm^{-1}$,
$d= 10^{-6}~cm$, $a=0.1$ at t=15 (in units of $t_0$).}
\label{Fig6}
\end{figure}

\begin{figure}
  \centering
  \includegraphics[width=70mm]{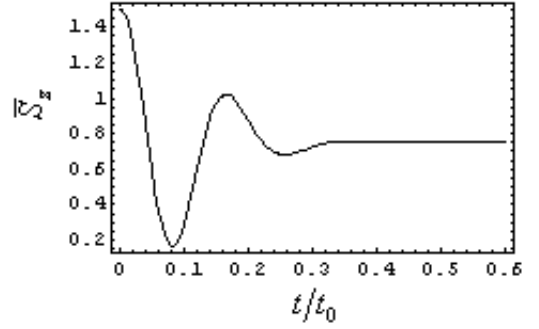}
\caption{(Color online) The average value of angular momentum
$\bar S_z(t)$ for initial spinor, Eq.(6) for $a=6$.}
\label{Fig7}
\end{figure}

For the second example presented in Sec.II, Eq.(17), the initial
values of effective spin are $\bar S_x(0)=\bar S_y(0)$, $\bar
S_z(0)=1/2$, and space-time evolution of spin density components
$S_i(\vec r,t)$ are similar to that discussed above. Note that for
these two cases the initial vector $\bar {\vec S}(0)$ is
perpendicular to the average momentum $\bar {\vec p}_0=(\hbar
k_0,0,0)$. But as was revealed in Ref.[19], the most interesting
spin dynamics characterized by hole spin precession occurs when
$\vec \rho (0)$ is not perpendicular or parallel to $\bar{\vec
p}_0$.

Such initial condition is realized in particular for the wave
packet given by Eq.(27). Using the expressions for the components
of the wave function at arbitrary time Eqs.(28), (29), (30), (31)
one can obtain the hole spin density $\vec S( \vec r,t)$ and the
components of average spin $\bar{\vec S}(t)$

$$\displaylines{\bar{S_x}(t)=-\frac{\sqrt{3}}{2}+\frac{\sqrt{3}}{8a^2}\Bigg[1-\exp(-\frac{a^2(t/t_0)^2}{1+(t/t_0)^2})\times\cr\hfill
\times\cos(\frac{a^2t/t_0}{1+(t/t_0)^2})\Bigg].\hfill\llap{(36)}\cr}$$

$$\displaylines{\bar{S_y}(t)=-\frac{\sqrt{3}}{2}\frac{\exp(-\frac{a^2(t/t_0)^2}{1+(t/t_0)^2})}{1+(t/t_0)^2}\Bigg(\sin(\frac{a^2t/t_0}{1+(t/t_0)^2})+
\cr\hfill
+\frac{t}{t_0}\cos(\frac{a^2t/t_0}{1+(t/t_0)^2})\Bigg)+\frac{\sqrt{3}}{4a^2}\exp(-\frac{a^2(t/t_0)^2}{1+(t/t_0)^2})\times\cr\hfill
\times \sin(\frac{a^2t/t_0}{1+(t/t_0)^2}).\hfill\llap{(37)}\cr}$$

$$\displaylines{\bar{S_z}(t)=\frac{1}{4}\Bigg[1+\frac{3\exp(-\frac{a^2(t/t_0)^2}{1+(t/t_0)^2})}{1+(t/t_0)^2}\times\cr\hfill
\times\Bigg(\cos(\frac{a^2t/t_0}{1+(t/t_0)^2})-
\frac{t}{t_0}\sin(\frac{a^2t/t_0}{1+(t/t_0)^2})\Bigg)\Bigg].\hfill\llap{(38)}\cr}$$

These expressions become more simple in the limit case
$d\rightarrow\infty$, i.e. for the plain wave propagating along
the $x$ axis with wave vector $k_0$

$$\bar S_x(t)=-\frac{\sqrt{3}}{2},~~~~\bar
S_y(t)=-\frac{\sqrt{3}}{2}\sin\frac{2\gamma_2\hbar k_0^2t}{m},$$

$$\bar S_z(t)=\frac{1}{4}\Bigg(1+3\cos(\frac{2\gamma_2\hbar
k_0^2t}{m})\Bigg).\eqno(39)$$

Last formulas clearly demonstrate the precession of vector
$\bar{\vec S}(t)$ about the wave vector $\vec k_0$: the end of
vector $\bar{\vec S}(t)$ describes ellipsis on the plane
$(S_y,S_z)$

$$\frac{(\bar S_y(t))^2}{m^2}+\frac{(\bar
S_z-1/4)^2}{n^2}=1,\eqno(40)$$ where $m=\sqrt{3}/2$, $n=3/4$.

In our case for the wave packet of finite width the time evolution
of the effective spin becomes more complicated as it follows from
Eqs.(36)-(38). Firstly, the component $\bar S_x(t)$ depends on
time, although this dependence is weak enough both for $a>>1$ and
$a<<1$. Secondly, the "precession" in $(S_y,S_z)$ plane acquires
the transient character which is demonstrated in Fig.7. As one can
see from Fig.7 the component $\bar S_y(0)=\bar S_y(\infty)=0$ and
the component $\bar S_z$ changes its values from $\bar S_z(0)=1$
to $\bar S_z(\infty)=1/4$.

To illustrate the spin density we plot the components $S_z(\vec
r,t)$ and $S_y(\vec r,t)$ in Fig.8 for the moments of the time
$t=0.1$ and $t=0.5$ (in units of $t_0$) correspondingly and for
$a=6$. The Fig. 8(b) demonstrates clearly the splitting of the
angular momentum density into two parts similar to probability
density discussed in Sec.II for $a>>1$. The angular momentum
density changes its sign within each part that looks like as
multipole structure.

\begin{figure}
  \centering
  \includegraphics[width=60mm]{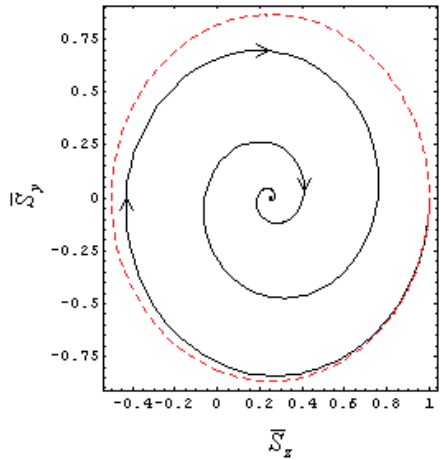}
\caption{(Color online) The transient precession of average
angular moment vector $\bar{\vec S}(t)$ (in plane $(S_y,S_z)$)
about the vector $\vec k_0~||~Ox$ for wave packet described by
Eq.(27) ($a=10$). Dashed line in accordance with Eq.(40)
corresponds to the hole spin precession for the plane wave.}
\label{Fig8}
\end{figure}

\begin{figure}
  \centering
  \includegraphics[width=70mm]{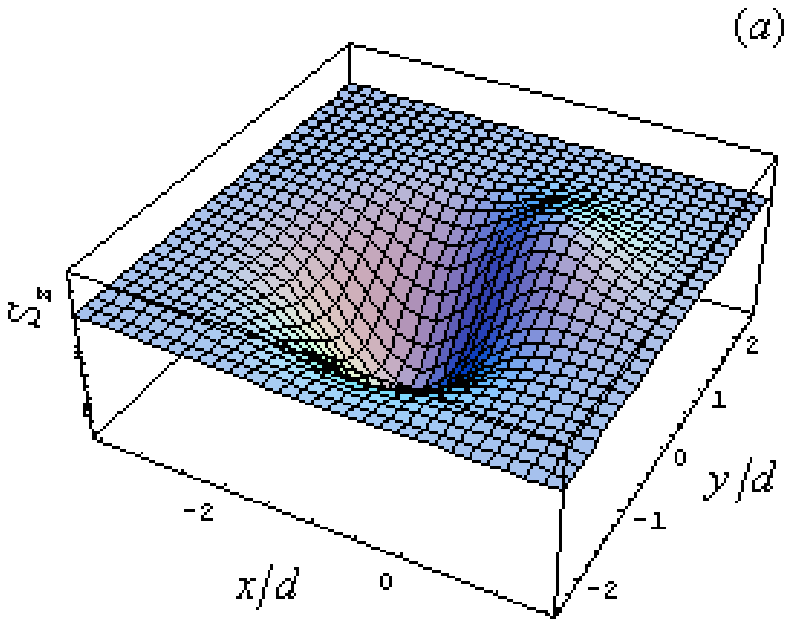}
  \includegraphics[width=70mm]{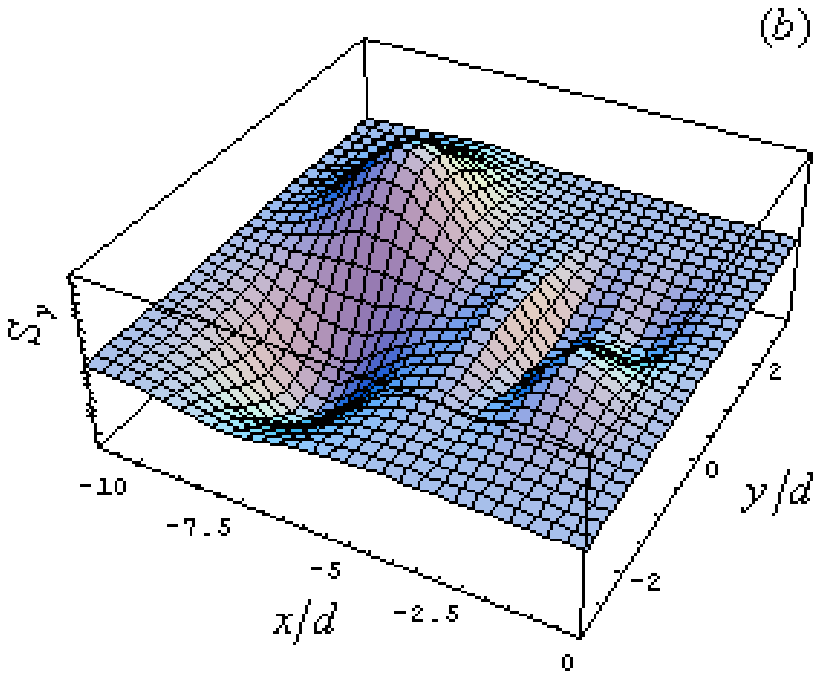}
\caption{(Color online) The angular momentum density $S_z(\vec
r,t)$ and $S_y(\vec r,t)$ for initial wave packet, Eq.(27) for
$k_0=2\cdot 10^{6}~cm^{-1}$, $d=3\cdot 10^{-6}~cm$, $a=6$ at
$t=0.1$ (a) and $t=0.5$ (b) (in units of $t_0$).} \label{Fig9}
\end{figure}

\section{Conclusion}

Using the Luttinger isotropic model we have studied the dynamics
of wave packets in hole 3/2 effective spin systems. The analytical
expressions for the four component wave function are obtained and
the probability densities as well as the effective spin densities
are visualized for different initial spin polarizations. Two
different types of the spatial evolution of wave packets are
found. In the case when the parameter $a>>1$ the initially
localized packet splits into two parts propagating with different
group velocities and having various effective spin polarizations.
Similarly to the wave packet evolution in systems with spin-orbit
coupling\cite{DMF} this evolution is accompanied by the
oscillations of packet center or {\it Zitterbewegung}. These
oscillations quickly fade away when the distance between two parts
of split packet exceeds its initial value. Another scenario of the
evolution is realized for a small value of parameter $a$, namely
when $a<<1$. In this case the probability density remains almost
axial symmetric with time, but the ripples appear at the
circumference of the density distribution. These ripples arise
owing to the interference of heavy- and light-hole states located
in the vicinity of the point $\vec k=0$ in momentum space. It
should be noted that this interference does not stimulate the
oscillations of packet center  $\bar{\vec r}(t)$ or ZB.

It is known that the spin dynamics of the holes, described by an
effective spin  - 3/2, differs qualitatively from that for spin  -
1/2 electrons. The main and striking peculiarity of hole spin
dynamics is the nontrivial periodic motion of average spin vector
in spin space, which can be viewed as a precession. We have
studied in detail such unusual behavior for the hole wave packets.
We have shown that finite width of wave packet is responsible for
the transient character of spin precession. The analytical and
numerical study of the influence of $a$ parameter on the angular
momentum density and average spin vector is presented. For the
case when $a>>1$ the density of angular momentum splits into two
parts having multipole polarization, but for $a<<1$ the angular
momentum density preserves almost the initial cylindrical
symmetry.

The experimental observation of trembling motion of $\vec r(t)$
and spin precession in crystalline solids is a long standing
problem of condensed matter physics. The possible methods of
producing of electron and hole wave packets as well as the
observation of ZB effects was analyzed in numerous of recent
publications. As was discussed in Ref. [20], the hole packets can
be excited optically in nondegenerate 3D hole gas by using a
polarized laser beam. If laser beam has a finite size the
optically exited wave packet will consist of states of light and
heavy holes lying in the range of $d^{-1}$ in vicinity of the
point $\vec k=0$ and can have a definite spin polarization. To
obtain the wishful average momentum $k_0$ the electric field pulse
with duration of some picosecond can be applied after excitation.
During the process of optical excitation electrons will excited
also, but the electric field will separate positive and negative
charges. The ZB evolution to be robust against disorder provided
$\omega_{ZB}>1/\tau_s$ and $\omega_{ZB}>1/\tau_p$ where $\tau_s$
and $\tau_p$ are spin and momentum relaxation times
correspondingly. As was verified experimentally (e.g. in $GaAs$
and $InAs$ $\tau_S$ is of the order $10-100$ ps) these conditions
can be fulfilled in clear semiconductors.

An interesting method of observation of trembling motion of
charged particles in solids was proposed by Rusin and Zavadski in
Ref. [15]. They note that the typical period of ZB oscillation is
comparable with the period of Bloch oscillations, and so the
method which early was used for observation of Bloch oscillation
in superlatticies at the presence of electric field\cite{Lyssenko}
can be used also for the ZB observation. The results of Sec. II
and Sec. III of this work show that this method can be used also
for ZB observation in Luttinger spin - 3/2 system.

The information about the evolution of hole wave packets is
important in the context of application for analysis of
functioning of different spintronic devices (e.g. Datta - Das hole
spin effect transistor\cite{Pala}) and for understanding the
transport phenomena and spin related dynamics.

\section*{Acknowledgments}
This work was supported by the program of the Russian Ministry of
Education and Science "Development of scientific potential of high
education" (project 2.1.1.2686), Grant Russian Foundation for
Basic Research (09-02-01241-a) and Grant of President of RF
MK-1652.2009.2.

\end{document}